# Modeling the Impact of Visual Brand Language on Attention, Object Recognition, and Memory Retrieval

Rachel F. Heaton and John E. Hummel

University of Illinois Urbana-Champaign
rfheaton@illinois.edu

**Abstract.** Visual brand language is the set of visual properties that convey brand identity for a product. What is the impact of visual brand language on a person's ability to recognize and understand the functional identity of an object? Using an empirically supported modeling framework based on the JIM model of object recognition and the LISA model of analogical inference, we simulated the impact of visual brand language on object recognition, the allocation of attention, and retrieval of functional information about objects. Our simulations predict that brand information captures attention and can slow recognition of an object's functional category, with greater degrees of branding causing larger effects. These results have potential implications for the usability and experience of designed objects.

## 1 Background

Brand identity is a key consideration in the design of consumer products, with the potential for both positive and negative effects. Branding communicates information beyond the mere function or category of a product, including information about the quality of manufacturing, the company's values, and its contextual expertise (e.g., a specialization in sports, beauty, or cooking products) [1]. Branding can create a positive association and bias a person toward choosing to purchase one product over another. Making the brand recognizable is therefore important if a company wishes to leverage prior successes into future sales [2]. On the other hand, strong branding might just as easily cause a person to have a negative association with a product, and excessive branding might create usability problems if the branding elements draw enough attention to interfere with a person's mental representation of an object while it is in use.

One of the primary ways that branding is communicated is via *visual brand language*. Visual brand language is the set of visual properties that are associated with the brand. Visual brand language is typically intentionally curated by designers, with explicit branding guidelines applied over different product lines. One of the most noticeable branding elements is the corporate logo, but color is another salient indicator of information about brand [3]. For example, tool manufacturer DeWalt uses yellow and black, agricultural machinery company John Deere uses green and yellow, and Nike sporting goods often feature orange, white, and black. Physical and tactile branding elements might include material, texture, and finish (e.g., Apple uses aluminum for the body of a laptop). Designers also strategically use typographic elements as part of brand identity: The typeface used by Nike (a brand-specific typeface based on the Futura typeface) is repeated in multiple contexts, and the "Just Do It." advertising slogan appears on Nike t-shirts, shoeboxes, and other products. Nike's broader visual brand language includes elements such as the color orange, "Just Do It." in the Nike typeface, and the logo swoosh.

Branding is, however, almost always independent of the functional identity of an object. An object with no brand information at all will be understood as being a hammer if it is shaped like a hammer. Each big box hardware store has a self-branded line of tools with their own color, and yet they all perform the equivalent functions. Yet branding could influence recognition of function: A hammer branded as a well-known tool manufacturer's product will likely be more recognizable as a tool. Conversely, a new product category that is completely inconsistent with the brand's previous line of products might be more difficult to recognize. A real-world example of a mismatch between branding and function is a coffee maker made by the tool brand Makita. Intuitively, the coffee maker's visual brand language activates a schema for power tools more than a schema for making coffee. However, whether the visual brand language of the coffee maker has a substantial impact on the functional interaction with the object is unclear.

## 2 Aims

Could branding consume enough cognitive resources to meaningfully interfere during an interaction with an object? To explore this question, we simulated the effects of visual brand language on object recognition using a modeling framework derived from the JIM model of object recognition [4,5,6] and the LISA model of relational reasoning [7,8, 9,10,11].

## 3 Simulating the Effect of Visual Brand Language on Object Recognition

Psychological models of object recognition have typically focused on one of two approaches, but to the best of our knowledge, neither approach has heretofore explored the effect of brand on object recognition and memory retrieval. One popular approach to modeling object recognition is a deep neural network approach where objects are recognized holistically on the basis of previously viewed examples [12, 13]. This approach to the theory of object recognition implicitly includes information that could be relevant to brand, like color and texture, but doesn't make such information explicitly available independent of the other properties of the object [11, 14,15]. Additionally, this approach provides no account of the inferences that people might make about the function of objects based only on their shape.

A second approach to modeling object recognition considers the relations among geometric volumes to be the basis for object recognition [4,5,6,16,17,18,19,20]. The parts-based approach has been successful in accounting for human experimental data in object recognition, including viewpoint invariance (see [21] for a review); the ability to recognize objects when parts have been replaced, removed, or added [16]; and inference for functional affordances [11]. A key property of this approach is that it represents elements of the object independently of each other and their relations. However, as outlined above, brand identity includes elements that may not be related to the physical, geometric structure of an object, and such elements have never been incorporated into these models.

In the work reported here, we augmented an existing structure-based model of object recognition [17] which was based on the upper layers of the JIM family of models [4,5,6]. The JIM models are based on Biederman's Recognition by Components theory of object recognition (RBC) [16], and represent objects as *structural descriptions* of geometric primitives, *geons*, into hierarchies of spatial relationships with each other. Geons are viewpoint independent categories of simple volumetric shapes, including cylinders, cones, rectangular prisms, etc. Geons are defined by properties like the shape of the cross section of the volume, the curvature of the axis, whether the sides are parallel or nonparallel, and the aspect ratio relative to the major axis. Although inverse optics is fundamentally ill-posed, the JIM model uses a heuristic algorithm to infer 3-D geon properties from non-accidental 2-D properties of the image [4].

The structural description of a coffee mug would be a geon with a curved cross section and a curved major axis (i.e., the handle) attached at two points to the side of a geon with a curved cross section and a straight major axis (the body of the mug). The same geons could be rearranged into different spatial relations, forming a bucket. The same neurons in each object represent cross section and major axis, regardless of whether they are encoding a coffee mug or a bucket. The binding of geons into spatial relationships is achieved via temporal synchrony of firing. For example, to represent that the handle is attached to the top of the bucket, the neurons representing the handle's properties fire in synchrony with the neurons representing the 'above' role of the relation, while the neurons representing the body of the bucket fire in synchrony with the 'below' role of that relation. Critically, the above *above*+handle neurons must fire out of synchrony with the *below*+body neurons. Training the model consists of a single exposure to each object, after which point the model can recognize the object in any viewpoint where most of the parts and relations are visible. JIM is not yet image computable but see [22] for effort in that direction.

Object recognition is only the first stage of interacting with an object. Successful interaction requires reasoning about the object based on prior experience. For example, one might reason about a novel coffee maker by analogy to familiar coffee makers. Fortunately, the upper layers of the JIM model are very similar to the representations used in the LISA (*Learning and Inference with Schemas and Analogies*) model of analogical reasoning [7,8,9,10,11]. Like JIM, LISA uses synchrony of firing to bind tokens (e.g., whole objects) into relations with each other. LISA accounts for hundreds of findings in the relational and analogical reasoning literature, including the details of analogical access, mapping, and inference, patterns of human development in relational reasoning [9], patterns of decline in normal aging and frontotemporal dementia [23], and the effect of emotional content on reasoning [24], among others.

Relational reasoning is necessarily a multistage process. Given a novel *target* problem to be solved, the first step is to retrieve a relevant *source* problem, or *schema*, from long term memory. Once retrieved, the target can be mapped to the source, and based on that mapping, the source can be used to make novel inferences about the target. For example, upon encountering a novel coffee maker, the parts and relations of the coffee maker (e.g., the filter basket above the carafe), will retrieve other coffee makers from long term memory and one's knowledge of those coffee makers will help one to infer how to interact with this novel coffee maker (see [11]). A natural consequence of mapping the target onto the source is the induction of a schema that captures what the target and source have in common and deemphasizes the properties unique to each [25,26].

For our current purposes, the crucial step in this process is the first one, memory retrieval. Specifically, how does the presentation of a target stimulus, for example a novel coffee maker with Nike brand elements, affect the system's ability to retrieve other coffee makers from long term memory to serve as a source for reasoning about the target?

### 3.1 Model Architecture

Heaton & Hummel [17] describe an architecture that integrates the parts-based representation of shape used by the JIM family of models with the representational format used by the LISA analogical inference engine. Using representations like those in the upper layers of JIM, Heaton & Hummel's model simulates the encoding and retrieval processes in LISA that precede analogical mapping and inference. The model encodes representations of presented objects on the fly, simultaneously matching them to representations in long term memory. The resulting model simulates Biederman's [16] demonstration that objects can be recognized even when their structural descriptions have been modified by adding, removing, or replacing parts. Heaton & Hummel showed that, like people, the model can recognize an object with changes to its structural description, but most readily recognizes exact matches to familiar objects. The model's ability to recognize objects based on structural descriptions, as well as its ability recognize objects despite changes to their structural descriptions, makes it an ideal starting point for the simulation work reported here.

The model encodes objects as *object files* [27] which are representations of an object's properties. As instantiated in this model, each object file consists of five layers of units representing semantic features, relational roles and their arguments (i.e., geons/parts), role bindings, propositions, and finally whole objects (**Fig. 1**).

Semantic units act as the input to the object file and carry information about the features of, and relations among, the geons composing an object. As in LISA, semantic units are divided into *role semantics* and *argument semantics*. In the case of the current model, the arguments of the relational roles are object parts (geons) and other visual characteristics such as branding elements. Role semantics feed role units, which encode information about relations such as *above*, *below, in-front*, *behind*, *right-of*, and *left-of*. Role units are reused within the same object file when a relation is repeated within the object's structural description (e.g., three geons stacked such that the first geon is above the second and the second is above the third would reuse the same *above* and *below* units). Part semantics feed *argument units*. Argument units carry information about geon properties such as the shape of the cross section, the shape of the major axis, whether the sides are parallel, its aspect ratio, and so forth. Argument units are reused when a geon stands in more than one relation to the other geons in an object (e.g., a central geon that has a relation to two other geons). If the same kind of geon appears more than once within an object (e.g., there are two different cylinders in an object), a separate argument unit is used to represent each. Keeping individual *tokens* of geons separate from types is managed by tagging each geon with location information: Geons with different location tags are necessarily separate entities and are treated as such. The location tags do not affect object recognition or encoding and serve only to individuate different parts of an object.

Role and argument units are bound together in long-term memory by *subproposition* units (SPs) [7]. For each role in a relation, an SP unit is used to bind an argument to that role. For example, in the coffee maker (**Fig. 1**), one unit binds the relational role *above* to the filter basket (*above*+filter_basket), while a second SP unit binds the relational role *below* to the carafe (*below*+carafe). SP units are connected to *proposition* units, which integrate multiple role bindings into complete relations. Finally, all the proposition units feed the group unit, which represents the object file as a whole.

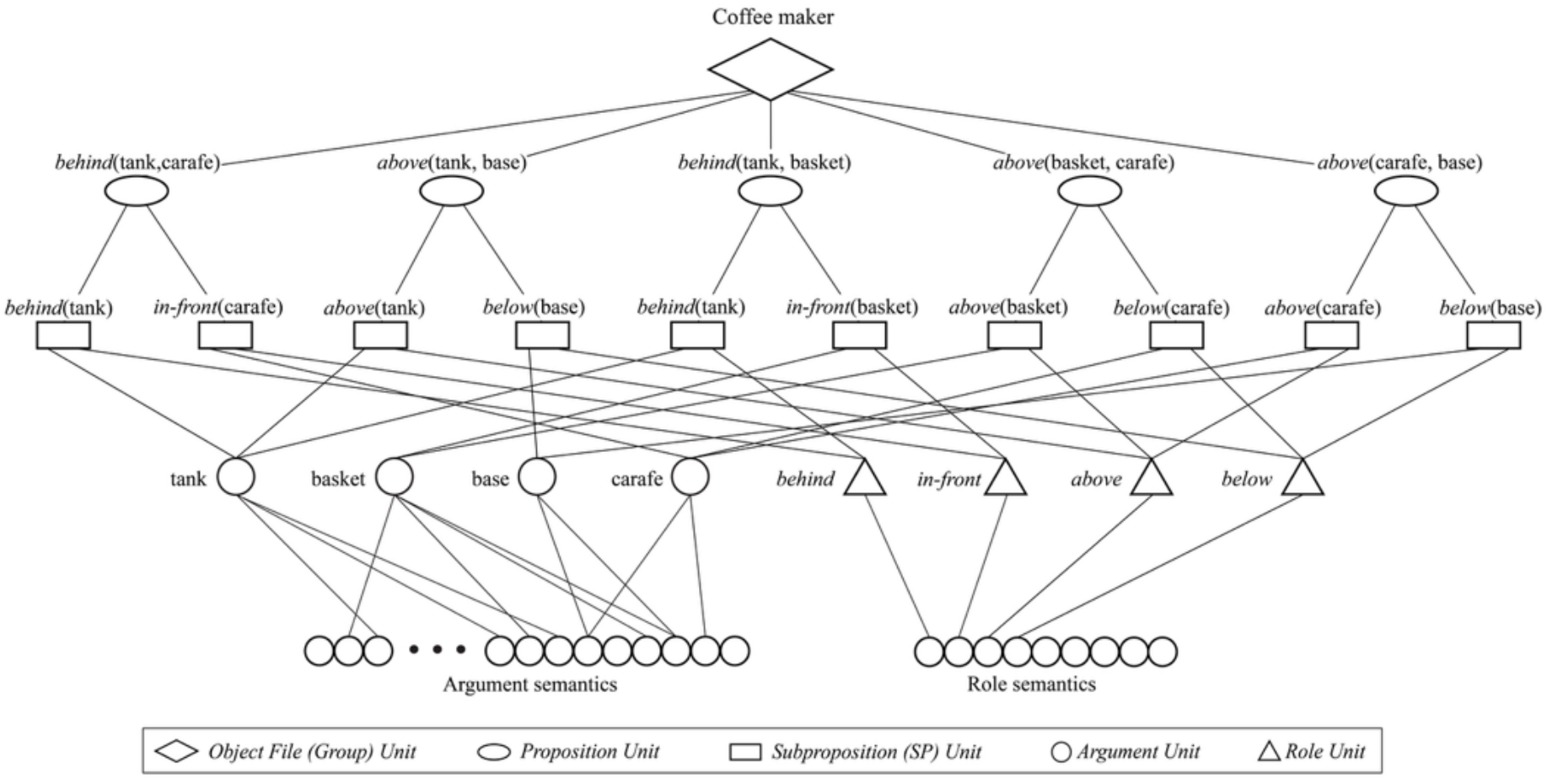

**Fig. 1.** Object file representation of a coffee maker.

Following Hummel & Holyoak [7,8], role and argument units, SP units, proposition units, and group units are implemented as leaky integrators that oscillate at different timescales. Role, part, and SP units oscillate at gamma (approximately 40Hz); proposition units oscillate at alpha (8-12Hz), and group units oscillate at theta (4-8Hz). At the end of each unit's oscillatory cycle, its activation is reset to zero, punctuating the cycle. Units of the same type within the object file laterally inhibit each other. The effect of both lateral inhibition and punctuation is that roles and arguments take turns being bound into propositions that correspond to the relations within the structural description, allowing the dynamic reuse of representations without binding errors (see [7]).

### 3.2 Recognizing and encoding objects into long term memory

Object files are the representational format in both the model's long-term memory and in perception, and the inputs to the model drive the activation of both existing object files in long term memory and the creation of a new object file in perception as the object is being presented. Each presentation of an object is encoded as a separate object file in long term memory. Objects are presented to the model as nested lists of roles and arguments composing propositions (**Table 1**). The construction of these input representations is a perceptual organization task performed by the early layers of JIM and is taken as given for the purposes of the current work [4].

In perception, when the object is first presented, a new group unit is recruited. At the end of every alpha cycle, a new proposition unit is recruited for each relation within the object. During the encoding of a proposition (at alpha), a new SP unit is recruited for each role binding in the relation (at gamma) and connected to the proposition unit. Role and argument (i.e., part) units are also connected to the SP. If no units currently exist to represent the role and argument being presented to the semantic units, new role and/or argument units are recruited. As the object file is being constructed, it is encoded into long term memory (see [17] for details).

**Table 1.** Objects composed of nested lists of propositions. The propositions describe relations between the object's parts.

| | |
|---|---|
| Coffee Maker | [[[above, carafe, location 1], [below, base, location 2]],<br>[[behind, water tank, location 3], [in front, carafe, location 1]],<br>[[above, basket, location 4], [below, carafe, location 1]],<br>[[behind, water tank, location 3], [in front, basket, location 4]],<br>[[above, tank, location 3], [below, base, location 2]]] |
| Nike T-Shirt | [[[orange color, shirt, location 5]],<br>[[left of, sleeve, location 13], [right of, shirt, location 5]],<br>[[left of, shirt, location 5], [right of, sleeve, location 14]],<br>[[print, swoosh, location 6], [substrate, shirt, location 5]],<br>[[print, Nike font, location 7], [substrate, shirt, location 5]]] |
| Nike Brand Identity | [[[orange color, generic object, location 8]],<br>[[print, swoosh, tag9], [substrate, generic object, location 8]],<br>[[print, Nike font, tag10], [substrate, generic object, location 8]]] |
| Subtle Nike Coffee Maker | [[[above, carafe, location 1], [below, base, location 2]],<br>[[behind, tank, location 3], [in front, carafe, location 1]],<br>[[above, basket, location 4], [below, carafe, location 1]],<br>[[print, swoosh, location 11], [substrate, basket, location 4]],<br>[[behind, tank, location 3], [in front, basket, location 4]],<br>[[above, tank, location 3], [below, base, location 2]]] |
| Overtly Branded Nike Coffee Maker | [[[orange color, base, location 2]],<br>[[orange color, basket, location 4]],<br>[[above, carafe, location 1], [below, base, location 2]],<br>[[behind, tank, location 3], [in front, carafe, location1]],<br>[[above, basket, location 4], [below, carafe, location1]],<br>[[behind, tank, location 3], [in front, basket, location 4]],<br>[[above, tank, location 3], [below, base, location 2]],<br>[[print, swoosh, location 11], [substrate, basket, location 4]],<br>[[print, Nike font, location 12], [substrate, base, location 2]]] |

Patterns of activation on the semantics activate role and argument units in existing object files, which activate SP and proposition units and eventually group units. An object file's group unit integrates its input at a timescale corresponding to theta oscillations in the brain. Within each theta cycle, one proposition (i.e., relation) is presented to the model at a time at a rate corresponding to alpha oscillations. Within each alpha oscillation, the individual role bindings are presented one at a time at a frequency corresponding to gamma.

The excitatory input, $e_i$, from semantic units, $j$, to argument (e.g., part) or role unit, $i$, is given by the Weber law:

$$e_i = \frac{\sum_j w_{ij} a_j}{\kappa + \sum_j w_{ij}}, \quad (1)$$

where $\kappa$, the Weber constant, is 10 for argument units (because of their relatively large number of excitatory inputs), and 1 for role and group units. The change in the activation of any unit $i$ at time $t$ is given by the leaky integrator:

$$\Delta a_i = \tau\gamma(1 - a_i)(e_i - i_i) - \delta a_i, \quad (2)$$

where $\tau$ is a time step constant ($\tau$ = 0.1 for argument, role, SP, and proposition units, and 0.01 for group units), $\gamma$ = 0.5 is a growth rate, $\delta$ = 0.1 is a decay rate, and $i_i$ is the inhibitory input to unit $i$:

$$i_i = \varepsilon * max_j(a_j), \quad (3)$$

where $\varepsilon$ is a constant and $j$ are other units of the same type as $i$ (e.g., argument, role, etc.) in the same object file as $i$.

The activation of SP units is given by equations 2 and 3, but at any given time *t*, the output of an SP is the max of its current activation and its current output. Although the activation of an SP oscillates at gamma along with the argument and role units feeding it, the output is sustained over time and oscillates at alpha, the timescale of proposition units. Similarly, although proposition unit inputs oscillate at alpha, the outputs oscillate at theta to feed group units.

## 4 Simulations

To explore the effect of branding on recognition, attention, and memory retrieval, we tested the model's ability to recognize coffee makers with different degrees of branding. We augmented JIM's strictly geon-based representations to include information about visual brand language. The library of arguments (i.e., object parts and brand elements) consists of 36-bit vectors that encode information about both 3-D and 2-D shape. The first 24 bits encode the properties of a geon, such as the shape of its cross section and major axis, whether its sides are parallel or nonparallel, whether it is pointed (like a cone) or truncated, and its aspect ratio. On a coffee maker, these bits would represent the shapes of parts such as its water tank, carafe, filter basket, and base. The last 12 bits encode 2-D information representing elements such as the Nike swoosh, Nike typography, and the parts of a flat t-shirt. There is also a semantically empty "generic object" consisting of 36 zeros. The library of relational roles consists of 9-bit vectors corresponding to roles like *above*, *below*, *behind*, *in-front*, *left-of*, *right-of*, *print* (that which is printed), and *substrate* (that which is printed on). The color orange is also represented as a role that can be bound to whatever part is that color.

Each simulation proceeded by encoding several objects in the model's long-term memory and then presenting one or more stimulus objects to observe the model's response to each stimulus (see **Table 2**). Every object consists of several propositions, which were presented to the model one at a time, in the order listed in **Table 1**. Because object files in memory compete with one another for retrieval, the order in which the propositions composing a stimulus are presented matters: Propositions presented first give some object files in memory a head start over others. Accordingly, we modeled the "salience" of visual properties (e.g., the Nike swoosh) in terms of the order in which they were presented, with more salient properties presented earlier.

### 4.1 Simulation 1

Simulation 1 presented the model with three versions of a coffee maker, one that was unbranded, one that was subtly Nike branded, and one that was overtly Nike branded (**Fig. 2**; see **Tables 1** & **2**). The unbranded version of the coffee maker only contained information about shape. The subtly branded version had the same shape but was augmented with a Nike swoosh printed on the carafe. The proposition representing the swoosh was presented late in the set of propositions, consistent with low visual salience. The overtly branded Nike coffee maker again had the same shape, but also contained two high salience branding elements and two low salience elements: The filter basket and the body of the coffee maker were orange, a swoosh was printed on the filter basket, and a Nike typographical element was printed on the body. The propositions representing the orange color were presented first, consistent with high visual salience. The swoosh and typographical element were presented last, consistent with lower visual salience.

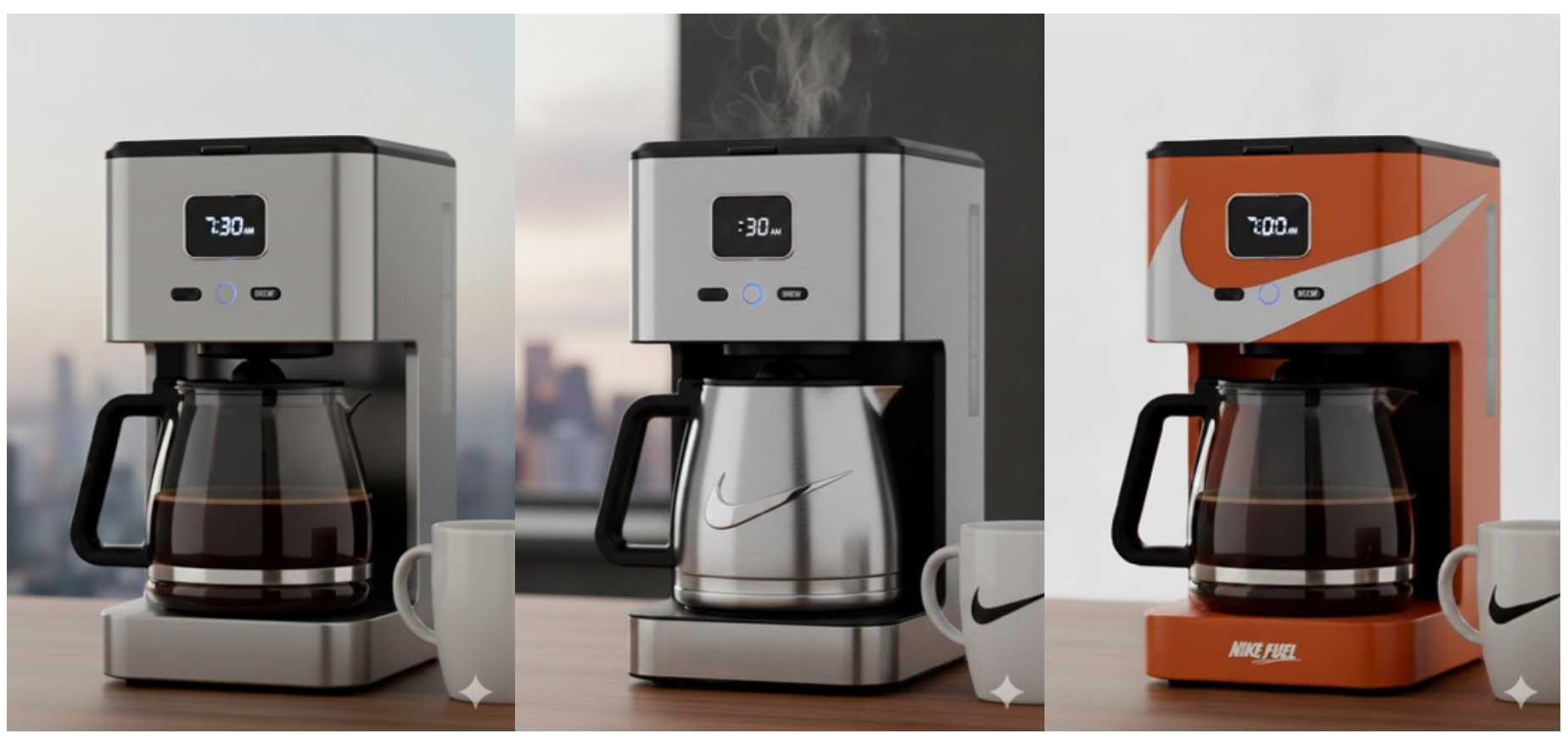

**Fig. 2.** Three versions of a coffee maker were presented to the model to test the effect of visual brand language on recognition and retrieval. Left: a coffee maker with no branding which matched the structural description of the object in long term memory. Center: a coffee maker with subtle Nike branding. Right: a coffee maker with overt Nike branding. Images were generated for illustration purposes using Google Gemini Nano Banana.

**Table 2.** Trials in Simulation 1.

| Trial | Stimulus | Long term memory | Retrieval target | Potential distractors |
|---|---|---|---|---|
| 1 | Unbranded coffee maker | Unbranded coffee maker, Nike t-shirt, Nike brand | Unbranded coffee maker | None |
| 2 | Subtly branded Nike coffee maker | Unbranded coffee maker, Nike t-shirt, Nike brand | Unbranded coffee maker | Nike t-shirt, Nike brand |
| 3 | Overtly branded Nike coffee maker | Unbranded coffee maker, Nike t-shirt, Nike brand | Unbranded coffee maker | Nike t-shirt, Nike brand |

Before each trial, the model was initialized with three representations in long term memory by encoding the unbranded coffee maker, the Nike t-shirt, and the Nike brand (**Table 2**). On the assumption that the coffee maker object file must be retrieved in order to reason by analogy about the branded coffee makers' ability to make coffee (see [11]), the activation of the unbranded coffee maker object file in long term memory was the test of recognition for the functional identity of the stimulus. The other two object files in long term memory both contained three Nike brand elements: the color orange, the swoosh, and Nike typography. The Nike t-shirt contained information about 2-D shapes in relations (a flat shirt with sleeves attached on the left and right sides) plus the three brand elements. The Nike brand identity object file represented the abstract visual brand language, and contained the same three brand elements, but contained no information about shape. Instead, the brand elements were attached to the "generic object", which had no semantic features.

We used two methods to evaluate the impact of branding on recognition, attention, and retrieval. The first measurement corresponded to recognition response time. We attached the object file group units to accumulators and measured the number of iterations it took the accumulators to cross a threshold of 100. On each iteration, the accumulator, $A_i$, for object file $i$ incremented its value by 0.1 times the activation, $a_j$, of the most active proposition, $j$, in $i$, and decremented its value by 0.002 times the most active proposition, $k$, in any group other than $i$, minus the activation of proposition $j$:

$$\Delta A_i = 0.1a_j - 0.002\left(a_k - a_j\right)^+, \qquad (4)$$

where the superscripted + indicates truncation below zero. The time it took the first group's accumulator to cross threshold was taken as the model's response time to classify the stimulus. We expected that all three versions of the coffee maker would eventually activate the unbranded coffee maker object file in memory, but the Nike branded coffee makers might also activate the Nike brand identity or the Nike t-shirt object files based on partial matches, slowing recognition of the functional identity of the stimulus (i.e., as a coffee maker).

The second index of the impact of branding was the proportion of attentional resources allocated to each object file on each iteration of the simulation, calculated as the activation of each group unit divided by the sum of the activations of all group units plus a baseline value of 0.05. To the extent that the Nike t-shirt and Nike brand identity become active with Nike branded coffee makers, we expected that attentional resources would be diverted away from the unbranded coffee maker object file toward the Nike object files.

## Results

In all three cases, the unbranded coffee maker object file crossed threshold first, indicating that the model recognized each coffee maker as a coffee maker. However, as expected, the unbranded coffee maker produced the fastest response time, driving the unbranded coffee maker object file past threshold in 3,063 iterations. For the Nike branded coffee makers, the prominence of the branding moderated the degree to which recognition was slowed down. In response to the subtly branded coffee maker, the unbranded coffee maker object file crossed threshold in 3,645 iterations, an approximately 19% percent slowdown relative to the unbranded stimulus. The overtly branded coffee maker was the most difficult for the model to recognize, with the unbranded coffee maker object file crossing threshold in 4,322 iterations, an approximately 41% percent slowdown. See **Fig. 3**.

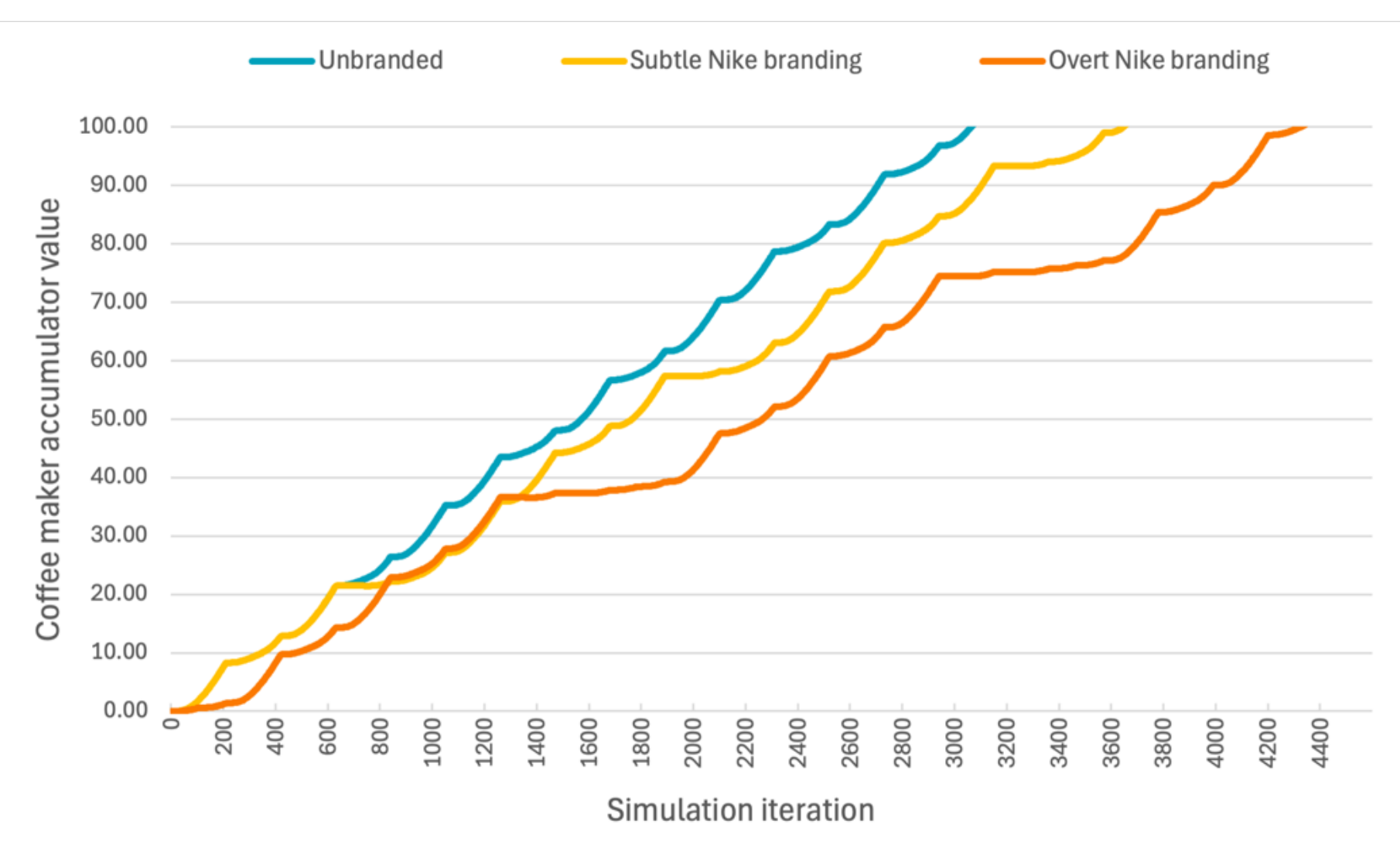

**Fig. 3.** Aggregate plot of the accumulator value of the unbranded coffee maker object in long term memory in response to the unbranded coffee maker (blue), the subtly Nike branded coffee maker (yellow), and the overtly Nike branded coffee maker (orange). On each trial, the unbranded coffee maker object file accumulator crossed threshold first. The other object files in memory are not shown. Recognition and retrieval were considered to occur when the accumulator crossed 100. The subtle branding has a moderate impact on recognition time, slowing recognition by 19%, but the overt branding results in a slowdown of 41%.

The time traces for the share of attention for each object file in memory (the unbranded coffee maker, the Nike brand, and the Nike t-shirt) after presentation of each Nike branded coffee maker are shown in **Fig. 4**. (The time trace for the presentation of the unbranded coffee maker is not shown because 100% of the attention was allocated to the unbranded coffee maker in memory for the entirety of the simulation.) In the case of the subtly branded coffee maker, all the model's attention was allocated to the unbranded coffee maker object file until late in the presentation, when the Nike branding captured about half the model's attention. When the overtly branded coffee maker was presented, the Nike brand and t-shirt object files captured attention early. By the middle of the simulation the coffee maker dominated attention, but by the end it shared attentional resources approximately equally with the Nike brand object files. These results suggest that branding has an impact on the allocation of attention, and overt branding may even dominate attention before an object's functional characteristics do.

Subtly branded Nike coffee maker
Unbranded coffee maker
Nike brand
Nike t-shirt
Proportion of total attention
1.0
0.9
0.8
0.7
0.6
0.5
0.4
0.3
0.2
0.1
0.0
0
200
400
600
800
1000
1200
1400
1600
1800
2000
2200
Simulation iteration

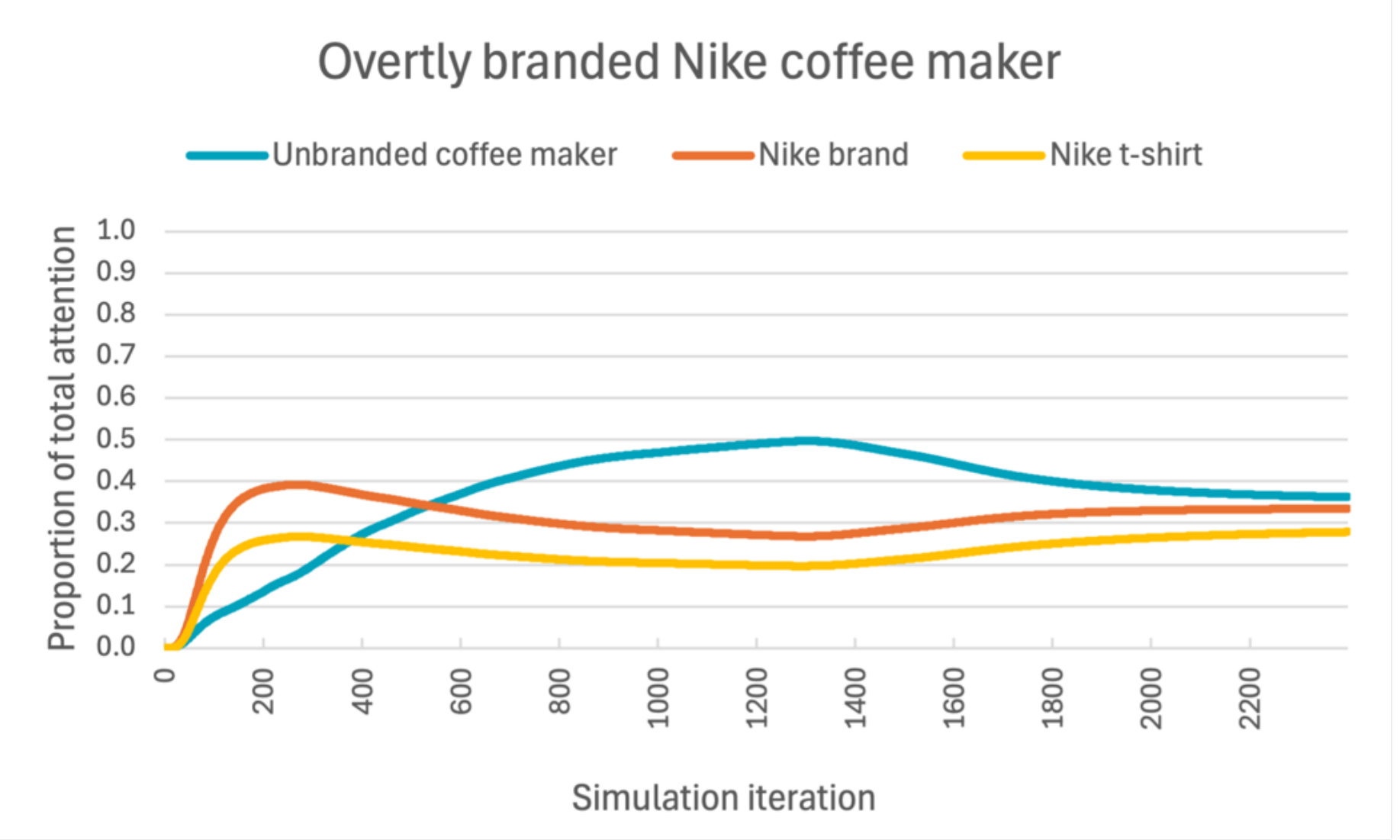

**Fig. 4.** Time traces for the share of attention of each object file in memory after presentation of the subtly Nike branded coffee maker (top) and the overtly Nike branded coffee maker (bottom). In response to the overtly Nike branded coffee maker, the Nike brand elements capture a greater proportion of the model's attention throughout the duration of the simulation.

Together, the results of Simulation 1 suggest that the degree to which an object is branded will modulate the viewer's ability to understand the functional characteristics of the object they are looking at. Overt branding captures a large proportion of attentional resources and slows recognition and retrieval.

### 4.2 Simulation 2

Simulation 2 explored whether encoding the overtly branded coffee maker into long term memory as an object that makes coffee mitigates the deleterious effects of branding. We initialized the model's long-term memory with the three original object files from Simulation 1 plus the overtly branded coffee maker, and then re-presented the overtly branded coffee maker as a stimulus for recognition. This time, on the assumption that the branded coffee maker could serve as a source analog for operating a coffee maker, we considered retrieval of either the unbranded coffee maker or the overtly branded coffee maker to be an index of recognition of the functional identity of the object (**Table 3**). As in Simulation 1, we used the time for the object file accumulators to cross threshold as an index of response time, and we recorded the time trace of the proportion of total activity of all the object files in long term memory as an index of attention.

**Table 3.** Stimulus and items in memory for Simulation 2.

| Stimulus | Long term memory | Retrieval target | Potential distractors |
|---|---|---|---|
| Overtly Nike branded coffee maker | Unbranded coffee maker, Nike t-shirt, Nike brand, Overtly Nike branded coffee maker | Unbranded coffee maker -or- Overtly Nike branded coffee maker | Nike t-shirt, Nike brand |

**Results**

Recall that in Simulation 1, the unbranded coffee maker was the stimulus that caused the unbranded coffee maker object file to cross threshold the fastest. In response to the branded coffee maker, the branded coffee maker in memory (purple line in **Fig. 5**) crossed threshold about as fast as the unbranded coffee maker did in response to the unbranded coffee maker stimulus in Simulation 1 (blue line in **Fig. 5**; 3,106 iterations in Simulation 2 vs 3,063 in Simulation 1). This implies that as long as the branded stimulus is encoded in memory, subsequent retrieval will be unharmed by the branding information. However, the unbranded coffee maker in memory (black line in **Fig. 5**) was still much slower to cross threshold in response to the branded stimulus in Simulation 2 (at 4,381 iterations) and in fact the response time was comparable to the time required to recognize the same stimulus in Simulation 1 (orange line in **Fig. 5**; 4322 iterations). In other words, having a branded coffee maker in long-term memory makes it faster to recognize that specific coffee maker, but it does not make it faster to recognize it as a generic coffee maker. This result suggests that any information about the original coffee maker that did not happen to migrate to the new Nike coffee maker in long term memory would take longer to infer due to a slowdown in retrieval of the original schema in response to the Nike coffee maker.

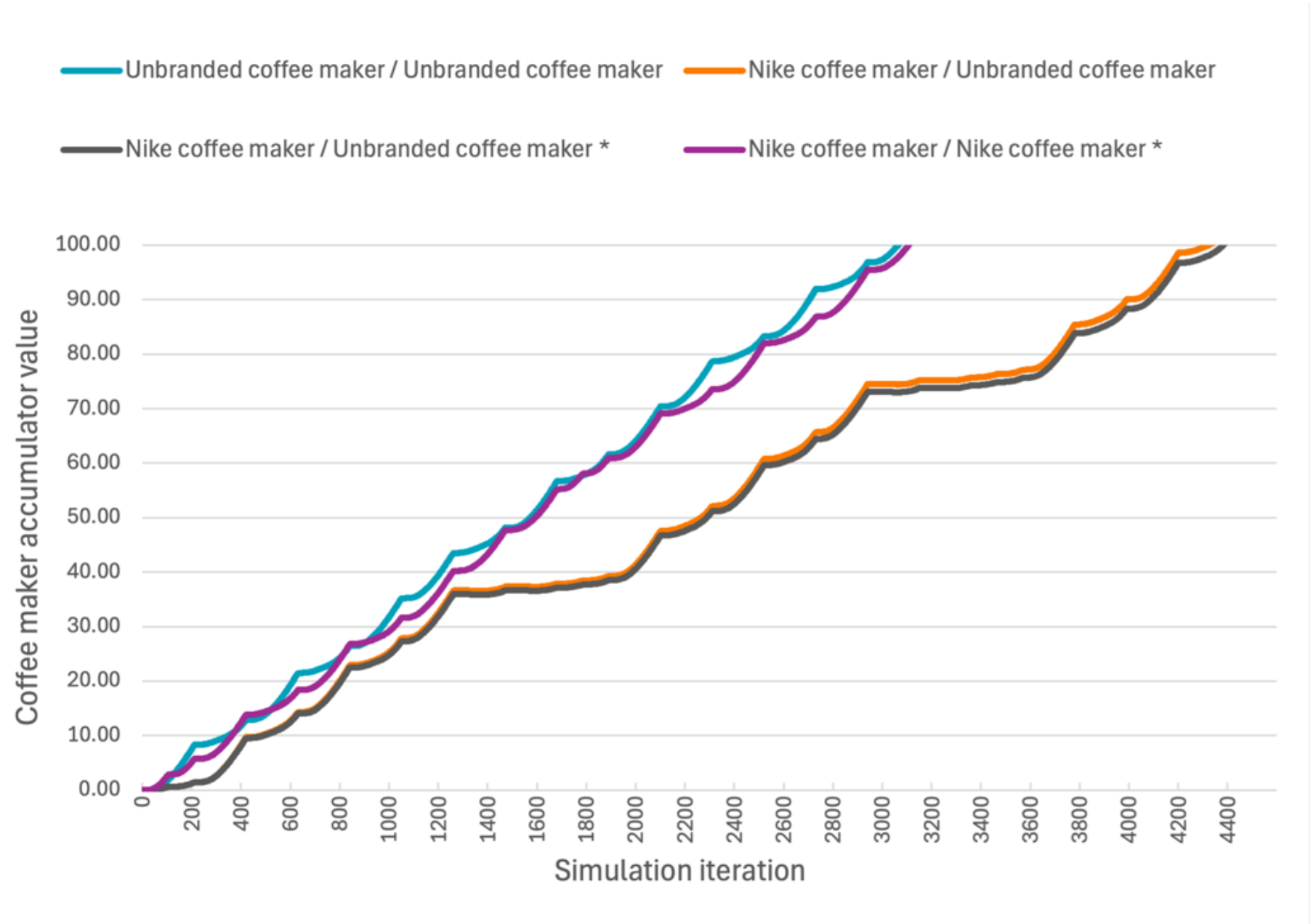

**Fig. 5.** Comparison of accumulators in Simulation 1 (blue and orange lines) and Simulation 2 (black and purple lines, indicated in the legend by *). In Simulation 1, the overtly branded coffee maker was not in long term memory, and the model responded most quickly to the unbranded coffee maker (blue line). In Simulation 2, the branded coffee maker was added to long term memory, and the branded coffee maker object file (purple line) was recognized faster than it did in Simulation 1 (compare purple and orange lines).

As can be seen in **Fig. 6**, early in the simulation, most of the model's attention was captured by the Nike brand and the Nike t-shirt. This result is unsurprising given that the orange color was the first proposition to be presented during the simulation. As the simulation progressed, the model's attention turned more to the object's shape, as evidenced by the attention devoted to the unbranded and overtly Nike coffee makers. By the end of the simulation, the model's attention was divided roughly equally among the four objects in its memory, suggesting that the overt Nike coffee maker's status as a coffee maker and its status as a Nike product captured approximately equal attention. Given that the Nike coffee maker is both a coffee maker and a Nike product, the case could also be made that, by the end of the simulation, roughly 75% of the model's attention was devoted to Nike in one way or another. Of course, 75% of the model's entire experience is also with Nike products, but even if the model had a realistic number of objects in its memory (e.g., bird houses, cats, mannequins, and socket wrenches), few of those would share features and relations with the overtly

Nike coffee maker. As such, 75% of what might realistically become active in response to a Nike coffee maker is perhaps not an unrealistic estimate. Interestingly, this 75% also includes the Nike t-shirt (roughly 20% of the model's attention at the end), even though t-shirts have nothing in common with coffee makers. It is not unreasonable to speculate, therefore, that upon seeing an overtly Nike coffee maker, one might be reminded of various specific Nike products.

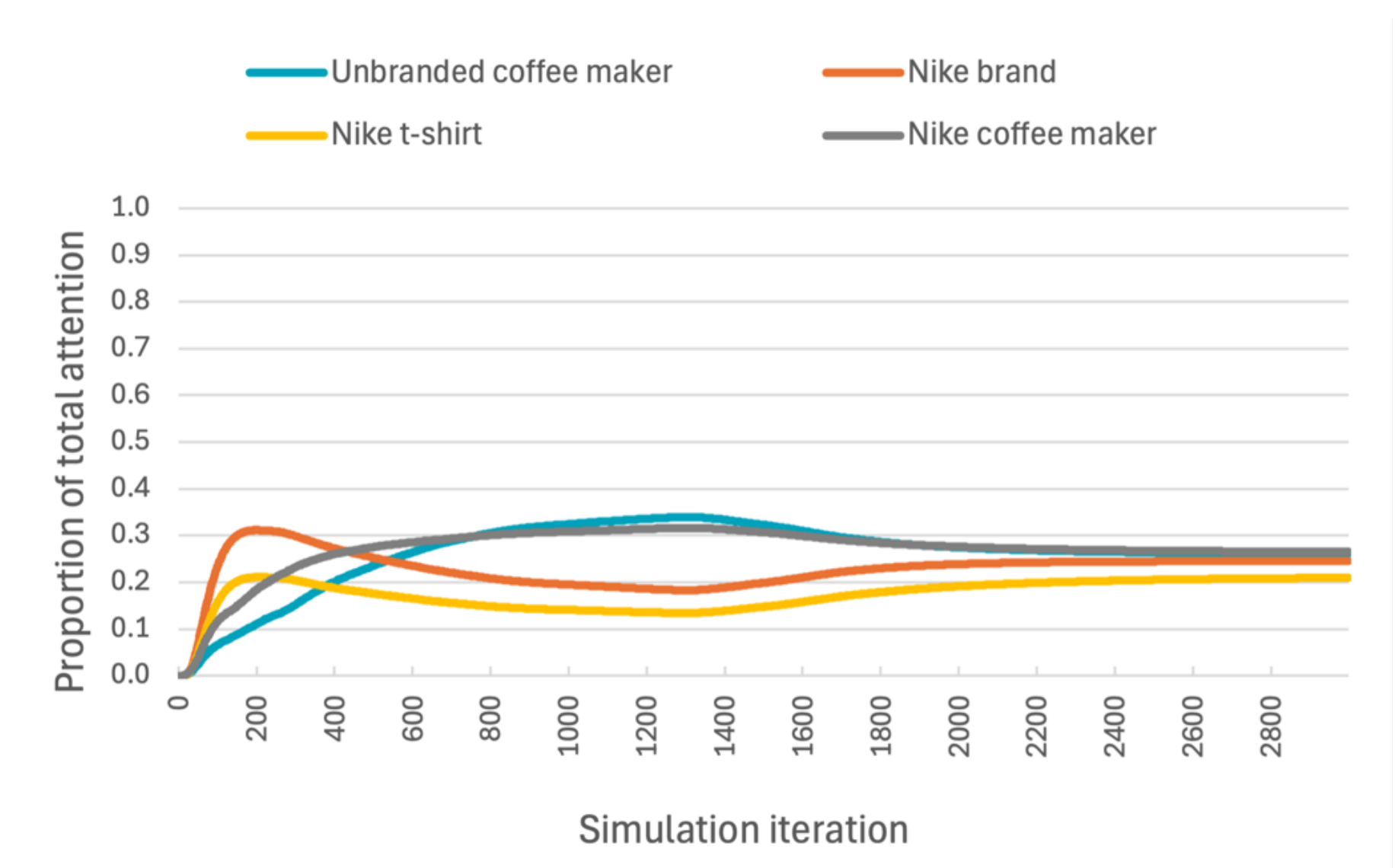

**Fig. 6.** Time traces for the share of attention of each object file in memory after the second presentation of the overtly Nike branded coffee maker. The Nike branded items, including the t-shirt, capture most of the model's attention early in the simulation. In the middle of the simulation, the shape properties capture a greater proportion of attention, and at the end the model attends equally to the Nike brand, the Nike coffee maker, and the unbranded coffee maker, with a small amount of attention still allocated to the t-shirt.

## 5 Discussion

Branding is helpful for consumers because they can use prior knowledge to make inferences about the experience they are likely to have with a product. If a consumer had a good experience with a particular brand of wrenches, they might look to reproduce that good experience when they need to buy a similar item, like a screwdriver [28]. Likewise, having a strong brand with positive associations is advantageous to companies because it can bias consumers toward repeated purchases from that same brand. This might be thought of as a kind of *halo effect*: positive associations in one domain can carry over to other domains [29,30]. Brand is an asset, often worth as much as the other tangible assets in the company, and brand itself can be monetized [31]. Clear and consistent visual brand language can be helpful to keep a brand and its products recognizable to consumers. When a product's functional identity is semantically consistent with the brand (e.g., sporting good products branded as Nike or combine harvesters branded as John Deere), the brand might even become an integral part of the schema for the product's functional category.

However, branding might have deleterious effects in some circumstances. Our simulations predict that even subtle branding has some impact on attention capture while using a product, and overt branding has an even greater impact. Usability is a continuum, with most product interactions falling somewhere between the total inability to perform a task and perfect efficiency. When confronted with an overtly Nike branded coffee maker, it is almost certain that people will figure out that the appliance makes coffee. The question is: How distracted will they be while they are making the coffee? Will the branding extend the time it takes to perform the task? Will they be unable to think about other things while making the coffee? Our results, using empirically supported models, predict that there is substantial distraction, not only upon first seeing the branded item and inferring its functional identity, but even after people have seen it before and have committed its functional identity to memory.

Distraction and delay can be more or less problematic, depending on the circumstances. The harm from overt branding on a defibrillator would be far more serious than overt branding on a coffee maker. However, if every interaction throughout the course of a day is subtly slowed by brand information, this could result in meaningful inefficiencies over the course of weeks, months, or years.

It is also possible that distraction from brand information could have sustainability implications by reducing a product's lifetime of use. Brand identity codes values and ethics, creating a potential point of psychological friction if the brand is damaged or the user's own values, ethics, and interests change. If a company adopts an unpopular political stance, the brand itself might become undesirable. Noticing the brand every time the item is used could become an irritant for the user. Even if the brand itself is not damaged, the user's own identity might change. An avid runner may love the idea of a Nike coffee maker, finding it motivating to get out the door for their morning training run. But if that person suffers an injury that ends their running career, the Nike brand identity displayed prominently on the kitchen counter might feel misaligned with their personal identity. To the extent that an object no longer reflects—or even conflicts with—the user's personal identity, the item might be prematurely discarded [32].

In the simulations reported here, we focused on just the first part of the process of visual reasoning about objects: attention, recognition, and retrieval. The part that we have simulated happens before explicit reasoning and after early visual processing. Our simulations show that the first effects of branding may right at the point of memory retrieval, with branding information processed before objects are even recognized.

There are potentially later effects of branding that are inevitable as the result of visual brand language. The schemas we used were limited to semantics describing the physical characteristics of the objects. However, the Nike brand identity might also contain other semantics and propositions related to sports (e.g., speed, agility, endurance, hard work, and determination) or the company itself (values, ethics). Hints of such effects were already apparent in our simulations when the Nike coffee maker activated the schema for the Nike t-shirt. As it happens, this type of remote reminding plays an important role in LISA's ability to make nonobvious inferences (see, e.g., [33]). Additional work using LISA could simulate additional inferences that could derive from branding beyond mere functional identity.

## 6 Conclusion

We investigated the effect of visual brand language on a computational model's ability to recognize, attend to, and retrieve information about objects from memory. Two simulations with an empirically well-supported model of human perception and reasoning showed that subtle brand language had a small effect on the model's attention to and recognition of a common consumer product (a coffee maker) whereas overt brand language had a much larger effect on the model's performance. These simulations suggest that branding likely impacts the allocation of attention, object recognition, and retrieval from memory, which in turn impacts the ability of a user to make inferences about an object. Future work includes testing these predictions, as well as additional simulations to explore other kinds of inferences that might be affected by brand language.